\documentclass[aps,prd,preprintnumbers,showpacs,showkeys,nofootinbib,superscriptaddress,fleqn,floatfix,tightenlines,10pt]{revtex4-2} 
\usepackage{amsmath,amsfonts,amssymb,amscd,amsxtra,amsthm}
\usepackage{dsfont}
\usepackage{graphicx}  
\usepackage{epstopdf}
\usepackage{dcolumn}  
\usepackage{bm}          
\usepackage{slashed}
\usepackage[utf8]{inputenc}
\usepackage{booktabs}
\usepackage[normalem]{ulem} 
\usepackage[dvipsnames]{xcolor} 
\usepackage{enumitem}
\usepackage{array}
\usepackage{slashed}
\usepackage{tikz}
\usepackage{float}
\usepackage{multirow}
\usepackage{booktabs} 
\usepackage{slashed}

\renewcommand\sout{\bgroup \color{red} \ULdepth=-.5ex \ULset}

\begin{document}
\preprint{INHA-NTG-03/2023}
\title{\Large Two-pole structure of the $b_1$(1235) axial-vector
  meson} 

\author{Samson Clymton}
\email[E-mail: ]{sclymton@inha.edu}
\affiliation{Department of Physics, Inha University,
Incheon 22212, Republic of Korea }

\author{Hyun-Chul Kim}
\email[E-mail: ]{hchkim@inha.ac.kr}
\affiliation{Department of Physics, Inha University,
Incheon 22212, Republic of Korea }
\affiliation{School of Physics, Korea Institute for Advanced Study 
  (KIAS), Seoul 02455, Republic of Korea}
\date{\today}
\begin{abstract}
We investigate the dynamical generation of the $b_1$ meson in the
$\pi\omega$ interaction, using the fully off-mass-shell
coupled-channel formalism with the $\pi\omega$, $\eta\rho$,
$\pi\phi$, and $K\bar{K}^*$ channels included. We first construct
the Feynman amplitudes for the sixteen different kernel amplitudes,
considering only the $t$ and $u$ channels. Solving the coupled 
integral equation, we obtain the transition amplitude for the
$\pi\omega$ interaction. We select the axial-vector and isovector
channels from the partial-wave expansion and single out the two poles 
corresponding to the $b_1$ mesons: $(1306-i70)$ MeV and $(1356-i65)$
MeV. They are located below the $K\bar{K}^*$ threshold. The first
pole lies below the $\eta\rho$ threshold by about 10 MeV, whereas the 
second one emerges above it by about 40 MeV. We analyze the effects of
the two poles and background contributions to the $\pi\omega$ total
cross section by using a toy model.  
\end{abstract}
\maketitle

\section{Introduction}
The low-lying pseudoscalar and vector meson
octets are believed to be $q\bar{q}$ ground states with quark spins
aligned antiparallel ($1{}^1S_0$) and parallel ($1{}^3S_1$),
respectively. The axial-vector mesons are also often considered as 
orbitally excited $q\bar{q}$ states. For example,   
there are six different light unflavored axial-vector mesons below 1.5
GeV: $h_1(1170)$, $b_1(1235)$, $a_1(1260)$, $f_1(1285)$, $h_1(1415)$, 
$f_1(1420)$~\cite{PDG}. Quark models classify $h_1(1170)$ and
$b_1(1235)$ as orbitally excited $1{}^1P_1$ states, and $a_1(1260)$ and
$f_1(1285)$ as $1{}^3P_1$ states~\cite{Godfrey:1985xj, Ebert:2009ub}.
However, since all of the axial-vector mesons lie above 1 GeV, a pair
of quarks can be created from the vacuum, so that the structure of
them may become more complicated. For instance, the $a_1(1260)$ meson
can be considered as a molecular 
state~\cite{Basdevant:1977ya, Lutz:2003fm, Roca:2005nm,
  Clymton:2022jmv}, which can be dynamically generated in $\pi\rho$
scattering within the coupled-channel formalisms.  
Since it appears as a resonance below the $K\bar{K}^*$ threshold, it
can be regarded as the $K\bar{K}^*$ molecular
state~\cite{Clymton:2022jmv}. Nagahiro et
al.~\cite{Nagahiro:2011jn} analyzed the compositeness of the $a_1$
meson and drew a conclusion that it has both the elementary and
composite components in comparable amounts. 

The $b_1(1235)$ meson can also be dynamically generated as
$a_1(1260)$~\cite{Lutz:2003fm, Roca:2005nm}. In
Refs.~\cite{Lutz:2003fm, Roca:2005nm}, it was shown that the
$b_1(1235)$ is strongly coupled to the $K\bar{K}^*$ channel. Its pole
position is located at
$(1288-i28)\,\mathrm{MeV}$~\cite{Roca:2005nm}. Experimentally, the 
$b_1(1235)$ was observed as a $\pi\omega$ resonance at around 1.22 GeV
in the $\pi^+ p\to \pi^+ + p + \pi^+ + \pi^- + \pi^0$
process~\cite{Abolins:1963zz}. Its existence was confirmed
consecutively~\cite{Karshon:1974qi, Chaloupka:1974eg,
  Gessaroli:1977mt, Evangelista:1980xe, OmegaPhoton:1983vde,
  OmegaPhoton:1984ols, Collick:1984dkp, Fukui:1990ki,
  IHEP-IISN-LANL-LAPP-KEK:1992puu, ASTERIX:1993wam}. In particular,
Fukui et al.~\cite{Fukui:1990ki} analyzed the $\pi\omega$ interaction
using the partial wave analysis, based on the experimental data on the
$\pi^- p$ charge exchange reaction. In the $1^{+-}$ wave, they
measured respectively the mass and width of the $b_1(1235)$ as
$M=(1236\pm 16)$ MeV and $\Gamma = (151\pm 31)$ MeV, fitting the
intensity of the $1^{+-}$ wave with the Breit-Wigner form. However,
some of the data are off from this form, which may indicate that the
structure of the $b_1(1235)$ is not explained by an ordinary $P$-wave 
$q\bar{q}$ state. 

In the present work, we investigate the $\pi\omega$ interaction within
the framework of a fully off-shell coupled-channel formalism. Since
the $b_1(1235)$ decays into $\pi\omega$ (seen), $\eta\rho$ (seen),
$K\bar{K}^*$ (seen), and $\pi\phi$ ($<1.5\,\%$), we consider these
decay channels in the current formalism. Constructing all possible
kernel amplitudes, we solve the coupled integral equations
to derive the transition amplitudes. Unexpectedly, we find two $b_1$
mesons: $(1306-i70)$ MeV and $(1356-i65)$ MeV. The partial-wave
($1^+(1^{+-})$) cross section for $\pi\omega$ scattering resembles
that of $\pi\Sigma$ scattering, where $\Lambda(1405)$ reveals a
two-pole structure~\cite{Oller:2000fj, Jido:2003cb}. Albaladejo et
al.~\cite{Albaladejo:2016lbb} found the two-pole structure of the
$D^*(2400)$ in the light pseudoscalar and $D$ meson interaction in the
coupled-channel formalism (see a recent review~\cite{Meissner:2020khl}
for detailed discussion on the two-pole structures). Moreover, there
is a peculiar fact concerning the $b_1$ meson. In contrast to all
other axial-vector mesons, there is no excited state for the $b_1$ 
meson. For example, we have two excited $a_1$ mesons such as $a_1(1420)$ 
and $a_1(1640)$, $f_1(1420)$ and $f_1(1510)$ for the $f_1$
meson, and $h_1(1415)$ and $h_1(1595)$ for the $h_1$ below 2 GeV. Even
in the strange channel, we have two excited states $K_1(1400)$ and
$K_1(1650)$ for the $K_1(1270)$~\cite{PDG}. Thus, the two-pole structure of the 
$b_1$ meson may give a clue in understanding the reason why the excited 
$b_1$ mesons have not been found.

The current work is organized as follows: In the next section, we
introduce the fully off-shell coupled-channel formalism. The 
transition kernel potentials are constructed by using the effective
Lagrangian. Incorporating them in the coupled integral equations, we
derive the transition amplitudes for the $\pi\omega$
interaction. In Sect.~\ref{sec:3}, We carry out the partial-wave
expansion to get the 
$1^+(1^{+-})$ channel corresponding to the $b_1$ meson. We compare the
total cross section for $\pi\omega$ scattering in the $1^+(1^{+-})$
channel with the data~\cite{Fukui:1990ki}. We pin down
the two poles for the $b_1$ in the $T$ matrix in the complex plane. We
construct a toy model including the two $b_1$ mesons to compare the
results with those from the current work and data. In the last
section, we summarize the current work and draw conclusions. 

\section{Coupled channel formalism\label{sec:2}} 
\begin{figure}[htbp]
	\centering
	\includegraphics[scale=1.2]{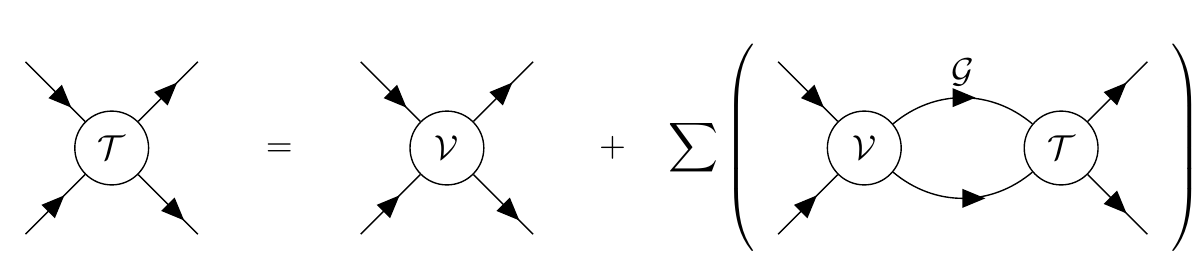}
	\caption{Graphical representation of the coupled integral
          scattering equation.}  
	\label{fig:1}
\end{figure}

The scattering amplitude is defined as 
\begin{align}
\mathcal{S}_{fi} = \delta_{fi} - i (2\pi)^4 \delta(P_f - P_i)
  \mathcal{T}_{fi}, 
\end{align}
where $P_i$ and $P_f$ stand for the total four momenta of the initial
and final states. The transition amplitude $\mathcal{T}_{fi}$ can be
derived from the Bethe-Salpeter integral equation 
\begin{align}
\mathcal{T}_{fi} (p',p;s) =\, \mathcal{V}_{fi}(p',p;s) 
+ \frac{1}{(2\pi)^4}\sum_k \int d^4q 
\mathcal{V}_{fk}(p',q;s)\mathcal{G}_{k}(q;s) \mathcal{T}_{ki}(q,p;s),  
\label{eq:2}
\end{align}
where $p$ and $p'$ denote the relative four-momentum of the
initial and final states, respectively. $q$ is the off-mass-shell
momentum for the intermediate states in the center of momentum
frame. $s$ represents the square of the total energy, which is just
one of the Mandelstam variables, $s=P_i^2=P_f^2$. The coupled integral
equation given in Eq.~\eqref{eq:2} can be depicted as in
Fig.~\ref{fig:1}.
To avoid the complexity due to the four-dimensional integral equation,
we make a three-dimensional reduction. While there are several
different methods for the three-dimensional reduction, we take the
Blankenbecler-Sugar scheme~\cite{Blankenbecler:1965gx, Aaron:1968aoz},
which takes the two-body propagator in the form of 
\begin{align}
	\mathcal{G}_k(q) =\;
  \delta\left(q_0-\frac{E_{k1}(\bm{q})-E_{k2}(\bm{q})}{2}\right)
	\frac{\pi}{E_{k1}(\bm{q})E_{k2}(\bm{q})}
  \frac{E_k(\bm{q})}{s-E_k^2(\bm{q})},  
\label{eq:4}
\end{align}
where $E_k$ represents the total on-mass-shell energy of the
intermediate state, $E_k = E_{k1}+E_{k2}$, and $\bm{q}$
denote the three-momenta of the intermediate state. Utilizing
Eq.~\eqref{eq:4}, we obtain the following coupled integral equation 
\begin{align}
	\mathcal{T}_{fi} (\bm{p}',\bm{p}) =\, \mathcal{V}_{fi}
	(\bm{p}',\bm{p}) 
	+\frac{1}{(2\pi)^3}\sum_k\int \frac{d^3q}{2E_{k1}(\bm{q})E_{k2}
	(\bm{q})} \mathcal{V}_{fk}(\bm{p}',\bm{q})\frac{E_k
	(\bm{q})}{s-E_k^2(\bm{q})} 
	\mathcal{T}_{ki}(\bm{q},\bm{p}),
	\label{eq:BS-3d}
\end{align}
where $\bm{p}$ and $\bm{p}'$ are the relative three-momenta
of the initial and final states in the center of momentum frame,
respectively.  

\begin{figure}[htbp]
	\centering
	\includegraphics[scale=0.5]{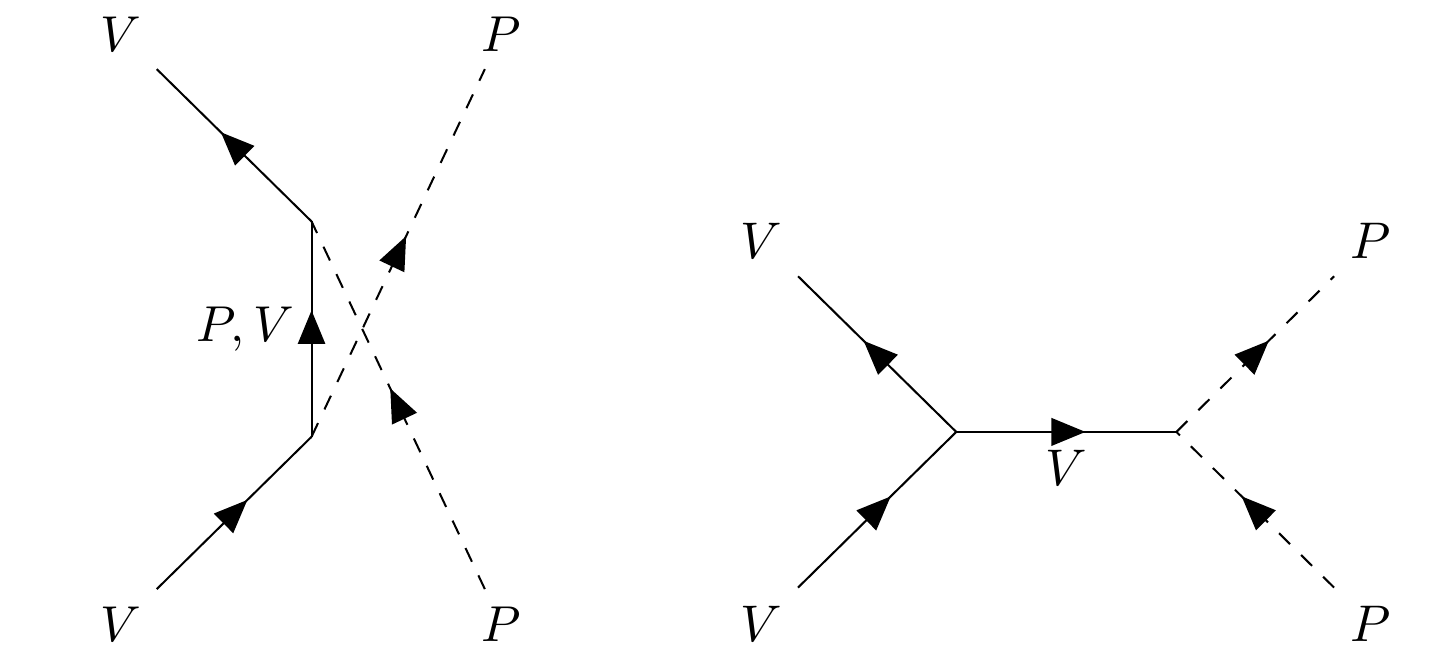}
	\caption{$u$- and $t$-channel diagrams for the
          meson-exchanged diagrams are depicted in the left and right
          panels, respectively. $P$ and $V$ stand for the
          pseudoscalar and vector mesons.} 
	\label{fig:2}
\end{figure}

We first construct the kernel amplitudes in Eq.~\eqref{eq:BS-3d} by
computing the tree-level Born diagrams, where the initial and final
states are given in terms of the pseudoscalar and vector mesons, as
illustrated in Fig.~\ref{fig:2}. Note that we do not include any 
pole diagrams in the $s$ channel.
The effective Lagrangians describe the interacting vertices
given in Fig.~\ref{fig:2}. They are given as   
\begin{align}
\mathcal{L}_{PPV} &= \sqrt{2}g_{PPV} \,\mathrm{Tr} 
		 \left([P,\partial_\mu P]\,V^\mu\right),\cr 
\mathcal{L}_{VVV} &= -\sqrt{2}g_{VVV} \,
		\mathrm{Tr}\left((\partial_\mu V_\nu
- \partial_\nu V_\mu)V^\mu V^\nu\right),\cr 
\mathcal{L}_{PVV} &=\sqrt{2}\frac{g_{PVV}}{m_V}\,\varepsilon^
	{\mu\nu\alpha\beta}\mathrm{Tr}\left(\partial_\mu
V_\nu\partial_\alpha V_\beta P\right). 
	\label{eq:su3sym}
\end{align}
Here, $P$ and $V$ denote the SU(3) matrices for the pseudoscalar and
vector meson octets, respectively
\begin{align}
P = \begin{pmatrix}
    \frac{1}{\sqrt{2}} \pi^0+\frac{1}{\sqrt{6}}\eta & \pi^+ & K^+\\
    \pi^- & -\frac{1}{\sqrt{2}} \pi^0+\frac{1}{\sqrt{6}}\eta & K^0\\
    K^- & \bar{K}^0 & -\frac{2}{\sqrt{6}}\eta
  \end{pmatrix},\;\; V = \begin{pmatrix}
    \frac{1}{\sqrt{2}} \rho^0_\mu+\frac{1}{\sqrt{2}}\omega_\mu &
    \rho_\mu^+ & K_\mu^{*+}\\
    \rho_\mu^- & -\frac{1}{\sqrt{2}} \rho_\mu^0+\frac{1}{\sqrt{2}}
    \omega_\mu & K_\mu^{*0} \\
    K_\mu^{*-} & \bar{K}^{*0}_\mu & \phi_\mu,
  \end{pmatrix}.
\end{align}
Note that we consider the ideal mixing for the vector nonet, while
treating $\eta_8$ as the physical $\eta$ meson for the  
pseudoscalar octet. We regard the $\rho$ meson as a gauge vector meson
in the hidden local symmetry~\cite{Bando:1984ej,Bando:1985rf}. Thus,
the coupling constants in the first two Lagrangians were identical,
while the last one was related to the anomaly term explaining the
$\omega$ decays into $\pi\rho$. We adopt the value of the coupling
constant from Ref.~\cite{Janssen:1994uf}, which is given as 
$g_{\pi\pi\rho}^2/4\pi=2.84$ and $(g_{\pi\rho\omega}^2/4\pi)
m_\omega^2 = 7.5$, which can be related to
the coupling constant in Eq.~\eqref{eq:su3sym} by the SU(3) symmetric
factor: $g_{\pi\pi\rho} = 2g_{PPV}$ and $g_{\pi\rho\omega}m_\omega =
2g_{PVV}$. We assume that the SU(3) symmetry is respected, except for
the $\phi$ coupling, which we allow to vary by approximately 1~\%.

The Born amplitudes in the $t$- and $u$-channels can be expressed in
terms of three different types of amplitudes:  
\begin{align}
	\label{eq:APu}
\mathcal{A}_{P}^{u}(\bm{p}',\bm{p}) =& -\mathrm{IS}\,g^2_{PPV}\,
 F^2(u)\,\left(2p_2-p_3\right)\cdot\epsilon^*
\mathcal{P}(p_1-p_4)\,\left(2p_4-p_1\right)\cdot\epsilon,\\
\label{eq:AVu}
\mathcal{A}_{V}^{u}(\bm{p}',\bm{p}) =& -\mathrm{IS}\,\frac{g^2_
{PVV}}{m_V^2}\,F^2(u)\,\varepsilon_{\mu\nu\alpha\beta}\,p_3^\mu\epsilon^
{*\nu}(p_3-p_2)^\alpha\,\mathcal{P}^{\beta\delta}(p_1-p_4)\,
\varepsilon_{\gamma\sigma\eta\delta}p_1^\gamma 
\epsilon^\sigma(p_1-p_4)^\eta,\\
\label{eq:AVt}
\mathcal{A}_{V}^{t}(\bm{p}',\bm{p}) =& -\mathrm{IS}\,g^2_{PPV}\,
F^2(t)\,\left(p_2+p_4\right)^\mu \mathcal{P}_{\mu\nu}(p_1-p_3)\left[(2p_1-p_3)\cdot\epsilon^*\epsilon^\nu+(2p_3-p_1)
\cdot\epsilon\epsilon^{*\nu}-\epsilon\cdot\epsilon^*(p_1+p_3)^\nu
  \right].  
\end{align}
The first two amplitudes represent pseudoscalar- and vector-meson
exchanges in the $u$-channel, while the last one represents
vector-meson exchange in the $t$-channel. The flavor components of 
Eq.~\eqref{eq:su3sym} can be evaluated in the isospin bases and yield
factors labeled as IS. The polarization vectors of the incoming and
outgoing vector particles are labeled as $\epsilon$ and $\epsilon^*$, 
respectively. The four-momenta of the particles involved are given as
follows: 
\begin{align}
	p_1 &= (E_{1,\mathrm{off}},\bm{p}),
	\hspace{0.3 cm} p_2 = (E_{2,\mathrm{off}},-\bm{p}),\cr
	p_3 &= (E_{3,\mathrm{off}},\bm{p}'),
	\hspace{0.3 cm} p_4 = (E_{4,\mathrm{off}},-\bm{p}').
\end{align}
Following the BbS scheme, the off-shell energies of the
particle are written~\cite{Aaron:1968aoz} as 
\begin{align}
E_{1,\mathrm{off}}&=\frac{1}{2} \left(\sqrt{s} +
               E_{1}(\bm{p}) - E_{2}(\bm{p})\right),\;\;\; 
E_{2,\mathrm{off}}=\frac{1}{2}\left(\sqrt{s} + E_{2}(\bm{p}) -
               E_{1}(\bm{p})\right),\nonumber\\ 
E_{3,\mathrm{off}}&=\frac{1}{2}\left(\sqrt{s} + E_{3}(\bm{p}') -
               E_{4}(\bm{p}')\right),\;\;\;
E_{4,\mathrm{off}}=\frac{1}{2}\left(\sqrt{s} + E_{4}(\bm{p}') -
               E_{3}(\bm{p}')\right), 
\end{align}
where they will satisfy the on-shell energy condition when  
$E_{1}(\bm{p})+E_{2}(\bm{p}) = E_{3}(\bm{p}') +
E_{4}(\bm{p}') = \sqrt{s}$. The propagators for the spin-0 and spin-1
mesons are expressed by 
\begin{align}
	\mathcal{P}(p) &= \frac{1}{p^2-m^2},\;\;\;
	\mathcal{P}_{\mu\nu}(p) = \frac{1}{p^2-m^2}
  \left(-g_{\mu\nu}+\frac{p_\mu p_\nu}{m^2}\right), 
\end{align}
where the $m$ are the corresponding exchange masses. As in the
previous study~\cite{Clymton:2022jmv}, the energy-dependent terms in the
denominator of the propagator are turned off. 

Since the hadrons have finite sizes, we introduce a form factor
at each vertex. We use the following parametrization for the form factors 
\begin{align}
	F(t) = \left(\frac{n\Lambda^2-m^2}
	{n\Lambda^2-t}
	\right)^n, \hspace{0.5 cm}
	F(u) = \left(\frac{n\Lambda^2-m^2}
	{n\Lambda^2-u} 
	\right)^n,
\label{eq:13}
\end{align}
where $n$ is determined by the power of the momentum in the vertex. While
the cut-off masses $\Lambda$ in Eq.~\eqref{eq:13} are free parameters,
we take a strategy to reduce the uncertainties associated with their
values. We add a fixed amount of ($500-700$) MeV to the corresponding 
mass of the exchange meson, motivated by the fact that a heavier
particle typically has a smaller size, as indicated in
Refs.~\cite{Kim:2018nqf,  Kim:2021xpp}. Consequently, we have set the
value of $\Lambda$ to be larger than the corresponding meson mass by
around ($500-700$) MeV. To fit the data, however, we choose a larger
value of the cut-off mass for some exchanged particles (see
Table~\ref{tab:1}). Moreover, for simplicity, we neglect the energy
and angular dependence of the form factors as done in
Ref.~\cite{Janssen:1994uf}.   

As mentioned in the introduction, the $b_1$ axial-vector meson has the
following decay channels: $\pi\omega$, $\pi\phi$, $\eta \rho$, and
$K\bar{K}^*$ with the positive $G$-parity. Thus, we consider these
four different channels in the the potential matrix $\mathcal{V}$ 
as follows:
\begin{align}
\mathcal{V} &= \begin{pmatrix}
\mathcal{V}_{\pi\omega\to\pi\omega} & \mathcal{V}_{\pi\phi\to\pi\omega} &
\mathcal{V}_{\eta\rho\to\pi\omega} & \mathcal{V}_{K\bar{K}^*\to\pi\omega}\\ 
\mathcal{V}_{\pi\omega\to\pi\phi} & \mathcal{V}_{\pi\phi\to\pi\phi} &
\mathcal{V}_{\eta\rho\to\pi\phi} & \mathcal{V}_{K\bar{K}^*\to\pi\phi}\\
\mathcal{V}_{\pi\omega\to\eta\rho} & \mathcal{V}_{\pi\phi\to\eta\rho} &
\mathcal{V}_{\eta\rho\to\eta\rho} & \mathcal{V}_{K\bar{K}^*\to\eta\rho}\\
\mathcal{V}_{\pi\omega\to K\bar{K}^*} & \mathcal{V}_{\pi\phi\to K\bar{K}^*} &
\mathcal{V}_{\eta\rho\to K\bar{K}^*} & \mathcal{V}_{K\bar{K}^*\to K\bar{K}^*}
\end{pmatrix}.
\label{eq:potfi}
\end{align}
Each component of the potential matrix can be constructed by computing
the tree-level Born diagrams with one-meson exchanges, which are
listed in Table~\ref{tab:1} along with the corresponding IS factors
and cutoff parameters required for the calculation of the potential, as
given in Eqs.\eqref{eq:APu}-\eqref{eq:AVt}.

\begin{table}[htbp]
	\caption{\label{tab:1}The isospin times SU(3) factor (IS) and cutoff 
	parameter ($\Lambda$) in MeV for possible exchange diagrams for each 
	reaction. $m$ is the exchange mass.}
	\begin{ruledtabular}
	\centering\begin{tabular}{lcccr}
	 \multirow{2}{*}{Reaction} & \multirow{2}{*}{Exchange} & 
	 \multirow{2}{*}{Type} & \multirow{2}{*}{IS} &
	 \multirow{2}{*}{$\Lambda-m$} \\
	 & & & &
	 \\\hline
	$\pi\omega \to \pi\omega$ & $\rho$ & $u$ & $4$ & 1200 \\
	$\pi\omega \to \eta\rho$ & $\omega$ & $u$ & $4/\sqrt{3}$ & 1000 \\
	$\pi\omega \to K\bar{K}^*$ & $K$ & $u$ & $\sqrt{2}$ & 600 \\
	                           & $K^*$ & $t$ & $\sqrt{2}$ & 1150 \\
	$\pi\phi \to K\bar{K}^*$ & $K$ & $u$ & $-2$ & 600 \\
	                        & $K^*$ & $t$ & $-2$ & 800 \\
	$\eta\rho \to \eta\rho$ & $\rho$ & $u$ & $4/3$ & 830 \\
	$\eta\rho \to K\bar{K}^*$ & $K$ & $u$ & $\sqrt{6}$ & 600 \\
	                        & $K^*$ & $t$ & $\sqrt{6}$ & 1050 \\
	$K\bar{K}^* \to K\bar{K}^*$ & $\rho$ & $t$ & $1$ & 600 \\
	                            & $\omega$ & $t$ & $-1$ & 600 \\
	                            & $\phi$ & $t$ & $-2$ & 1400 \\
	$K\bar{K}^* \to \bar{K}K^*$ & $\pi$ & $u$ & $1$ & 600 \\
	                            & $\eta$ & $u$ & $-3$ & 600 \\
	                            & $\rho$ & $u$ & $-1$ & 600 \\
	                            & $\omega$ & $u$ & $1$ & 600 \\
	                            & $\phi$ & $u$ & $2$ & 1400 \\
	\end{tabular}
    \end{ruledtabular}
\end{table}

\begin{figure}[htbp]
	\centering
	\includegraphics[scale=0.6]{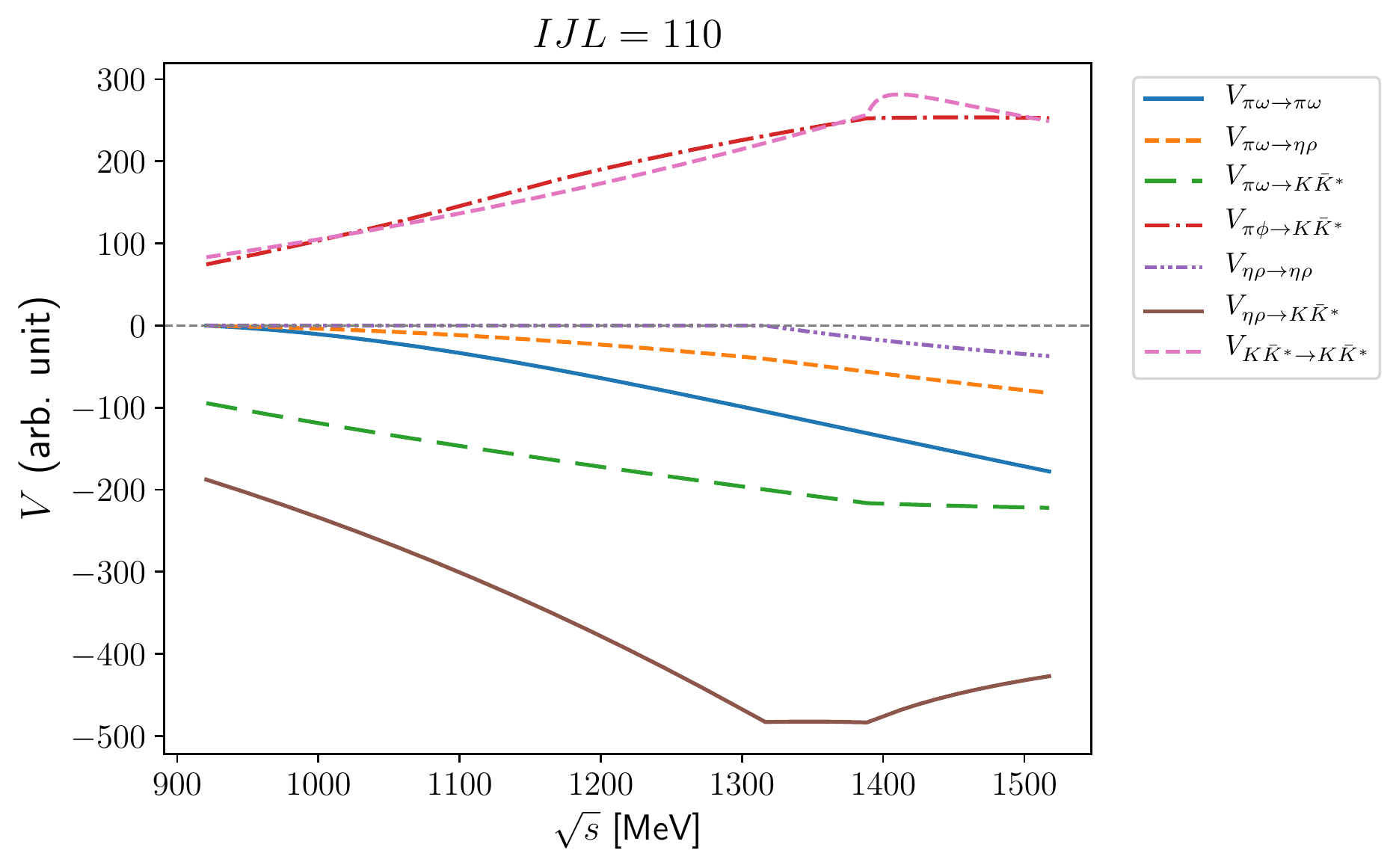}
	\caption{Numerical results for the transition potentials in
          seven different channels as functions of the total energy.}
\label{fig:3}
\end{figure}

In Fig.~\ref{fig:3}, we show the numerical results for the on-shell potentials. 
Note that channels without a tree-level amplitude have been omitted
from Fig.~\ref{fig:3}. Analysis of Fig.~\ref{fig:3} and Table~\ref{tab:1} 
reveals that the $\mathcal{V}_{\pi\omega\to\eta\rho}$ potential, which contains 
a single $u$ exchange diagram, is significantly smaller than the 
$\mathcal{V}_{\pi\omega\to K\bar{K}^*}$ potential, which contains $u$- and 
$t$- exchange diagram, even though the former has an IS factor more than 
twice as large as the latter. This discrepancy suggests that the $t$-exchange 
diagram is considerably stronger than the $u$-exchange diagram. Consequently, 
we observe in Fig.~\ref{fig:3} that all channels with $K\bar{K}^*$ as the 
initial or final state produce larger values of the potential due to
the presence of the $t$-exchange diagram. In particular, the  
$\mathcal{V}_{\eta\rho\to K\bar{K}^*}$ has the strongest contribution due to 
its large IS factor. Although $\mathcal{V}_{K\bar{K}^*\to K\bar{K}^*}$ has the 
three $t$-exchange diagrams, the $\rho$ and $\omega$ exchanges cancel each 
other because of the different sign of the IS factor, leaving only the $\phi$ 
exchange contribution, which a has smaller IS factor than that of
$K^*$ exchange in the $\eta\rho\to K\bar{K}^*$ process. These results
clearly indicate that the $K\bar{K}^*$ channel is crucial in
generating the resonances dynamically within the current
framework. Similar conclusions are found in Refs.~\cite{Lutz:2003fm,
  Roca:2005nm}. 

To focus on the $b_1$ meson, we need to carry out the partial-wave
expansion of the $\mathcal{V}$ and $\mathcal{T}$ matrices, resulting
in a one-dimensional integral equation given by 
\begin{align}
	\mathcal{T}^{J(fi)}_{\lambda'\lambda} (\mathrm{p}',\mathrm{p}) = 
	\mathcal{V}^{J(fi)}_{
	\lambda'\lambda} (\mathrm{p}',\mathrm{p}) + \frac{1}{(2\pi)^3}
	\sum_{k,\lambda_k}\int
	\frac{\mathrm{q}^2d\mathrm{q}}{2E_{k1}E_{k2}}
	\mathcal{V}^{J(fk)}_{\lambda'\lambda_k}(\mathrm{p}',
	\mathrm{q})\frac{E_k}{
	s-E_k^2} \mathcal{T}^{J(ki)}_{\lambda_k\lambda}
	(\mathrm{q},\mathrm{p}),
	\label{eq:BS-1d}
\end{align}
where $\lambda'$, $\lambda$ and $\lambda_k$ denote the helicities of 
the final ($f$), initial ($i$) and intermediate ($k$) states, 
respectively. The partial-wave $\mathcal{V}$ amplitudes can be
expressed as 
\begin{equation}
	\mathcal{V}^{J(fi)}_{\lambda'\lambda}(\mathrm{p}',\mathrm{p}) = 
	2\pi \int d(
	\cos\theta) \,d^{J}_{\lambda'\lambda}(\theta)\,\mathcal{V}^{fi}_{
	\lambda'\lambda}(\mathrm{p}',\mathrm{p},\theta),
\end{equation} 
where $\theta$ denote the scattering angle and $d^{J}_{\lambda'\lambda}
(\theta)$ stand for the matrix elements of the Wigner $D$ functions in the helicity basis. The partial-wave $\mathcal{T}$ amplitudes are expressed in a similar way.

The integral equation presented in Eq.~\eqref{eq:BS-1d} contains the 
singularity originating from the two-body meson propagator $\mathcal{G}$. 
To regulate this singularity, we isolate the singular part of the propagator. 
The resulting regularized integral equation is given by
\begin{align}
	\mathcal{T}^{fi}_{\lambda'\lambda} (\mathrm{p}',\mathrm{p}) = 
	\mathcal{V}^{fi}_{
	\lambda'\lambda} (\mathrm{p}',\mathrm{p}) + \frac{1}{(2\pi)^3}
	\sum_{k,\lambda_k}\left[\int_0^{\infty}d\mathrm{q}
	\frac{\mathrm{q}E_k}{E_{k1}E_{k2}}\frac{\mathcal{F}(\mathrm{q})
	-\mathcal{F}(\tilde{\mathrm{q}}_k)}{s-E_k^2}+ \frac{1}{2\sqrt{s}}
	\left(\ln\left|\frac{\sqrt{s}-E_k^{\mathrm{thr}}}{\sqrt{s}
	+E_k^{\mathrm{thr}}}\right|-i\pi\right)\mathcal{F}
	(\tilde{\mathrm{q}}_k)\right],\cr
	\label{eq:BS-1d-reg}
\end{align}
with 
\begin{align}
	\mathcal{F}(\mathrm{q})=\frac{1}{2}\mathrm{q}\,
	\mathcal{V}^{fk}_{\lambda'\lambda_k}(\mathrm{p}',
	\mathrm{q})\mathcal{T}^{ki}_{\lambda_k\lambda}(\mathrm{q},\mathrm{p}).
\end{align}
and $\tilde{\mathrm{q}}_k$ is the momentum when $E_{k1}+E_{k2}=\sqrt{s}$. 
The regularization is applied only when the total energy $\sqrt{s}$ exceeds 
the threshold energy of the $k$-th channel $E_k^{\mathrm{thr}}$. It is worth 
noting that the form factor that was introduced in the $\mathcal{V}$ 
amplitudes provides sufficient suppression in the high-momentum region, thus 
allowing for the regularization of the integration.

To obtain $\mathcal{T}$ from Eq.~\eqref{eq:BS-1d-reg} numerically, we 
expand the $\mathcal{V}$ matrix in the helicity states and momentum space, 
where the momenta are obtained by using the Gaussian quadrature method. We 
then derive the $\mathcal{T}$ matrix by using the Haftel-Tabakin method for 
the matrix inversion~\cite{Haftel:1970zz}, namely 
\begin{align}
	\mathcal{T} = \left(1-\mathcal{V}\tilde{\mathcal{G}}\right)^{-1} 
	\mathcal{V}.
\end{align}
Note that the $\mathcal{T}$ matrix is still in the helicity basis. To 
facilitate the analysis, we express the $\mathcal{T}$ matrix in the $IJL$ 
particle basis~\cite{Machleidt:1987hj}. The relations between the $\mathcal{T}$ 
matrix elements in the two bases are given by
\begin{align}
	\mathcal{T}^J_{J,J} &=\mathcal{T}^J_{++}-\mathcal{T}^J_{+-},\\
	\mathcal{T}^J_{J-1,J-1} &=\frac{1}{\sqrt{2J+1}}
	\left[J \mathcal{T}^J_{00}+(J+1)(\mathcal{T}^J_{++}+\mathcal{T}^J_{+-})
	+\sqrt{2J(J+1)}(\mathcal{T}^J_{+0}+\mathcal{T}^J_{0+})\right],\\
	\mathcal{T}^J_{J+1,J+1} &= \frac{1}{\sqrt{2J+1}}\left[(J+1) 
	\mathcal{T}^J_{00}+ J(\mathcal{T}^J_{++}+
	\mathcal{T}^J_{+-})-\sqrt{2J(J+1)}(\mathcal{T}^J_{+0}
	+\mathcal{T}^J_{0+})\right],
\end{align}
where we have only shown the diagonal part of $\mathcal{T}^J_{L',L}$ as it 
is relevant to the particle production. Here, we define $\mathcal{T}_{IJL}$ 
to be the component of the $\mathcal{T}$ matrix in the particle basis for a 
given total isospin $I$, total angular momentum $J$, and orbital angular 
momentum $L$.

\section{Results and discussions \label{sec:3}}
\subsection{The $b_1$ meson}
The existence of the $b_1$ meson is supported by experimental observations 
in pion-proton scattering~\cite{Fukui:1990ki,
IHEP-IISN-LANL-LAPP-KEK:1992puu}, proton-antiproton 
annihilation~\cite{ASTERIX:1993wam}, and 
photoproduction~\cite{OmegaPhoton:1984ols}. However, these hadronic
reactions are known to be sensitive to hadronic backgrounds, which 
causes significant uncertainties into the data. Thus, as an
alternative process, the $\tau$ decay was considered as a pristine one for
studying the axial vector meson. In contrast to the $a_1$ meson, however,
the $b_1$ meson in the $\tau$ decay has not yet been conclusively
identified. Efforts have been made to study the $b_1$ meson in the
$\tau\to \pi\omega \nu_\tau$ decay, but the dominance of the
first-class $\rho$-meson vector current in the decay makes it
difficult to discern the $b_1$ meson~\cite{CLEO:1999heg}. Thus, one
has to rely on hadronic reactions to examine the properties of the  
$b_1$ meson. In the current work, we will consider $\pi\omega$ scattering
arising from the charge-exchange $\pi\rho\to \omega \pi n$ reaction to
investigate the $b_1$ meson, specifically taking the experimental data
\cite{Fukui:1990ki} into account. 

\begin{figure}[htbp]
 \centering
 \includegraphics[scale=0.55]{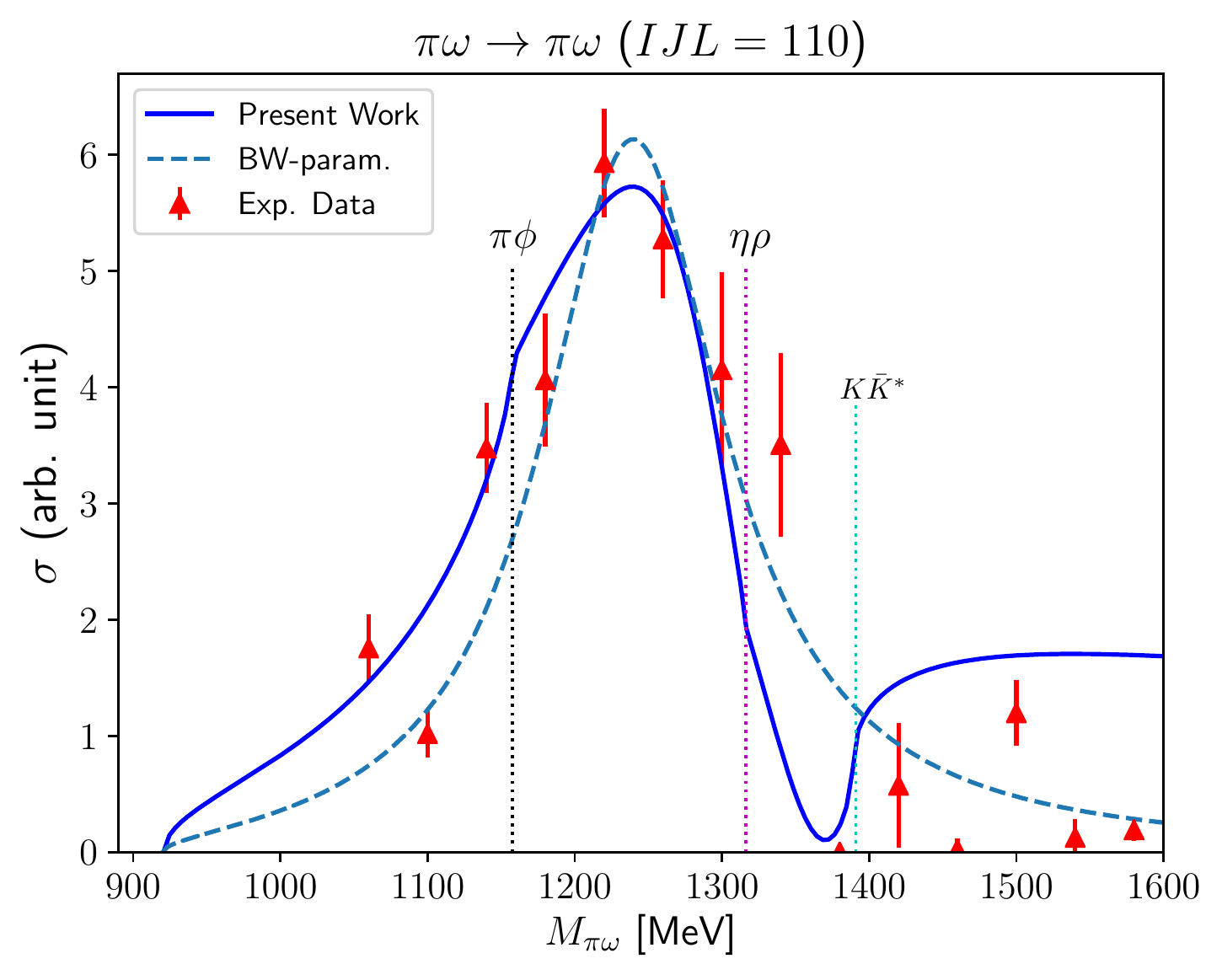}
 \caption{The $\pi\omega\to\pi\omega$ total cross
   section for $IJL=110$ as a function of $\pi\omega$ invariant
   mass. The experimental data are taken from
   Ref.~\cite{Fukui:1990ki}.}  
 \label{fig:4}
\end{figure}
We conduct the numerical computation of the invariant mass
distribution for the charge-exchange reaction, comparing the current
results with the experimental data~\cite{Fukui:1990ki}. We adjust the
cutoff masses as little as possible, while keeping all other
parameters the same as those determined in the previous study of 
the $a_1(1260)$ meson~\cite{Clymton:2022jmv}. To describe the
experimental data, we assume that the total cross section in arbitrary
unit is proportional to the $\pi\omega$ elastic $\mathcal{T}$
amplitude with a constant factor $C$, expressed as 
\begin{align}
	\sigma \equiv -C\, \mathrm{Im}[\mathcal{T}_{\pi\omega}(M_{\pi\omega})].
\end{align}
The resonance structure is generally sensitive to the cutoff
masses. It is understandable, given the fact that the singularities
generated dynamically in the current work arise from the off-shell
contribution to the $\mathcal{T}$ matrix. In the course of the fitting
process, we keep the cutoff masses to be $\Lambda = m+600$ MeV as far
as possible, where $m$ is the mass of the exchange particle. By doing
that we can suppress numerical uncertainties. However, we have found
that $\Lambda=1.4$ GeV should be used for the $\phi$-exchange diagram
(see Table~\ref{tab:1}). The results are in good agreement with the
experimental data, as demonstrated in Fig.~\ref{fig:4}. We observe
that the total cross section falls off drastically till the
$K\bar{K}^*$ threshold, which is a typical threshold behavior.  
Interestingly, the experimental data also exhibits a similar
tendency. We compare the present result with the single resonance
structure coming from the Breit-Wigner form~\cite{Fukui:1990ki}
\begin{align}
	\sigma = \rho(M_{\pi\omega}) \left| \frac{g^2}{M_{\pi\omega}^2
	-M^2+iM\Gamma}\right|^2,
\end{align}
where the mass and width are taken to be experimental values $M=1236$
MeV and $\Gamma = 151$ MeV, respectively.  The phase space factor is
given by $\rho=|\bm{p}|/8\pi M_{\pi\omega}$, where $\bm{p}$ is the
three-momentum of the particle in the CM frame.  
The BW form produces the experimental data near the resonance region
well, whereas it cannot explain the threshold behavior near the
$K\bar{K}^*$ threshold, which implies that the coupled-channel effects
are required. 

In order to investigate the properties of the $b_1$(1235) meson, we
search for the singularities in the $\mathcal{T}$ matrix in the
complex energy plane. The resulting pole position of 
$\sqrt{s_R}=(1306-i70)$ MeV, which corresponds to the width of 140
MeV, is consistent with both experimental data~\cite{Fukui:1990ki} and the 
average value provided by the PDG. In contrast, recent lattice 
calculations~\cite{Woss:2019hse} yielded a pole position for the $b_1$ meson 
at $[(1382\pm 15)-i (46\pm 15)]$ MeV, where the $\pi\omega$ and $\pi\phi$ 
scattering channels were coupled. Meanwhile, the chiral unitary 
model~\cite{Roca:2005nm} obtained a pole position at $(1247-i28)$ MeV for 
the $b_1$ meson. Both the works underestimate the width of the $b_1$ 
meson.

\subsection{Two-pole structure of $b_1$ meson}
\begin{figure}[htbp]
    \centering
    \includegraphics[scale=0.55]{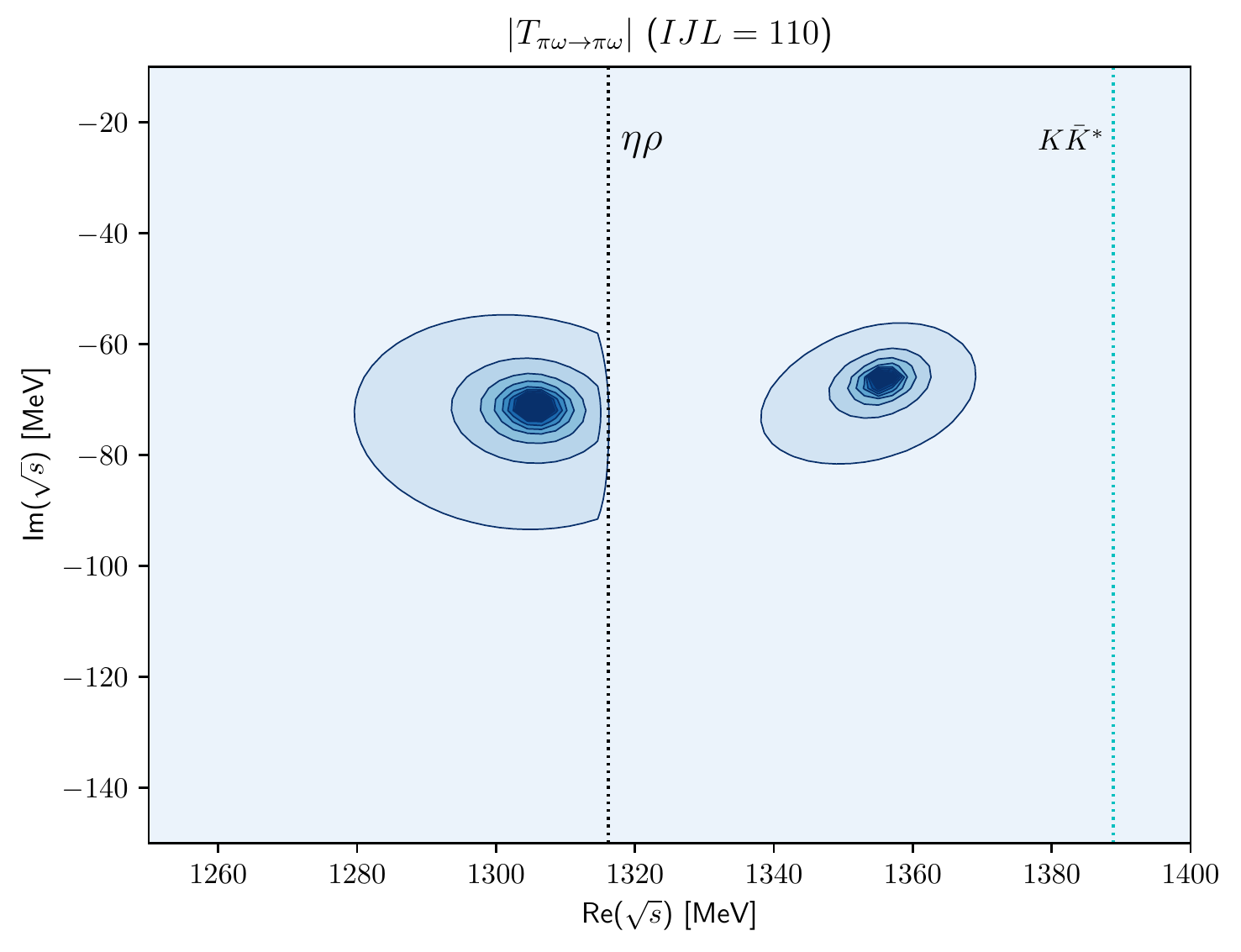}
    \caption{The $\pi\omega\to\pi\omega$ total cross
      section for $IJL=110$ as a function of $\pi\omega$ invariant
      mass. The experimental data are taken from
      Ref.~\cite{Fukui:1990ki}.}  
    \label{fig:5}
\end{figure}

As mentioned previously, we have searched for the singularities of the 
$\mathcal{T}$ matrix in the complex energy plane. In addition to the first 
pole $\sqrt{s_R}= (1306-i70)$ MeV, we also find the second pole at 
$(1356-i65)$ MeV. To show these two poles, we present a contour plot of 
the $\mathcal{T}_{\pi\omega \to \pi\omega}(IJL = 110)$ amplitude as a 
function of the complex energy in the second Riemann sheet. 
Figure~\ref{fig:5} displays the distinct two poles. The first pole appears 
below the $\eta\rho$ threshold, whereas the second pole comes out between 
the $\eta\rho$ and $K\bar{K}^*$ thresholds. Specifically, the first pole 
lies below the $\eta\rho$ threshold by about 10 MeV, while the  
second pole emerges above it by about 40 MeV and below the $K\bar{K}^*$ 
threshold by about 30 MeV. The widths of the two $b_1$ mesons are 
comparable each other.

To better understand the properties of the $b_1$ mesons, it is of great 
use to calculate the coupling strengths of the $b_1(1235)$ resonances, 
which can be derived from the residues of the $\mathcal{T}$ matrix:
\begin{align}
	\mathcal{R}_{a,b} &= \lim_{s\to s_R} \left(s-s_R\right)
	\mathcal{T}_{a,b}/4\pi,\\
	g_{a} &= \sqrt{\mathcal{R}_{a,a}},
\end{align}
Note that, since we define the coupling strengths in this way, their signs 
cannot be determined. The results for the coupling strengths are 
listed in Table~\ref{tab:2}. As shown in it, the $S$-wave 
$K\bar{K}^*$ channel is the most strongly coupled to both the poles: 
$g_{K\bar{K}^*} = (8.64 + i0.34)$ GeV and $g_{K\bar{K}^*} = (8.30 +
i2.78)$ GeV for the first and second poles,
respectively. Interestingly, the first pole for the $b_1$ meson has a
larger value of the $\eta\rho$ coupling strength  
than the second one. It is understandable, since the first pole is closer 
to the $\eta\rho$ threshold, compared to the second one. The first pole 
is also more strongly coupled to the $\pi\omega$ and $\pi\phi$ channels. 
However, we want to mention that it is essential to include all
channels so that the $b_1$ resonances can be generated dynamically. It
is also interesting to compute the ratios of the $D$- to $S$-wave
amplitudes in decays of the $b_1$ mesons to each channel. In
Table~\ref{tab:3}, we list the results for them. In particular, we
derive the value of 0.36 for the first pole, and 0.40 for the second
pole. The experimental data on the $D/S$ ratio is given as  
$0.277 \pm 0.027$. Since we have two poles for the $b_1$ meson, we
cannot directly compare the results with the experimental value. The
$D/S$ ratios for all other channels are generally very small. 
\begin{table}[htbp]
	\caption{\label{tab:2}Coupling strength in unit of GeV of the
          $b_1$ resonance to the $S$ and $D$ wave states in the four
          different channels.} 
	\begin{ruledtabular}
	\begin{tabular}{lcr}
	$\sqrt{s_R}$ [MeV] & $1306-i70$   & $1356-i65$\\
	\hline
    $g_{\pi\omega}$($S$-wave)    & $3.07+i2.14$ & $0.68+i2.91$\\ 
    $g_{\pi\omega}$($D$-wave)    & $1.33+i0.29$ & $0.84+i0.87$\\ 
    $g_{\pi\phi}$($S$-wave)      & $2.34+i1.35$ & $0.84+i1.84$\\
    $g_{\pi\phi}$($D$-wave)      & $0.20-i0.03$ & $0.19+i0.11$\\
    $g_{\eta\rho}$($S$-wave)     & $3.36+i1.91$ & $0.54+i3.16$\\ 
    $g_{\eta\rho}$($D$-wave)     & $0.03-i0.08$ & $0.48-i0.06$\\ 
	$g_{K\bar{K}^*}$($S$-wave)   & $8.64+i0.34$ & $8.30+i2.78$\\
	$g_{K\bar{K}^*}$($D$-wave)   & $0.18+i0.17$ & $0.02+i0.17$\\
	\end{tabular}
    \end{ruledtabular}
\end{table}
\begin{table}[htbp]
	\caption{\label{tab:3}$D/S$ amplitude ratios of the $b_1$ resonance 
	in the four different decay channels, obtained from the ratio of the 
	absolute value of the coupling strength of the $b_1$ to the $D$ and 
	$S$ wave states.}
	\begin{ruledtabular}
	\begin{tabular}{lcccc}
		\multirow{2}{*}{Pole} & \multicolumn{4}{c}{Channels}\\
		& $\pi\omega$ & $\pi\phi$ & $\eta\rho$ & $K\bar{K}^*$ \\
		\hline
		$1306-i70$ MeV& $0.36$ & $0.07$ & $0.02$ & $0.03$ \\
		$1356-i65$ MeV& $0.40$ & $0.11$ & $0.15$ & $0.02$
	\end{tabular}
    \end{ruledtabular}
\end{table}
The $D/S$ ratios for the 
second pole in the decays of $\pi\phi$ and $\eta\rho$ are about 10~\%.

\begin{figure}[htbp]
 \centering
 \includegraphics[scale=0.55]{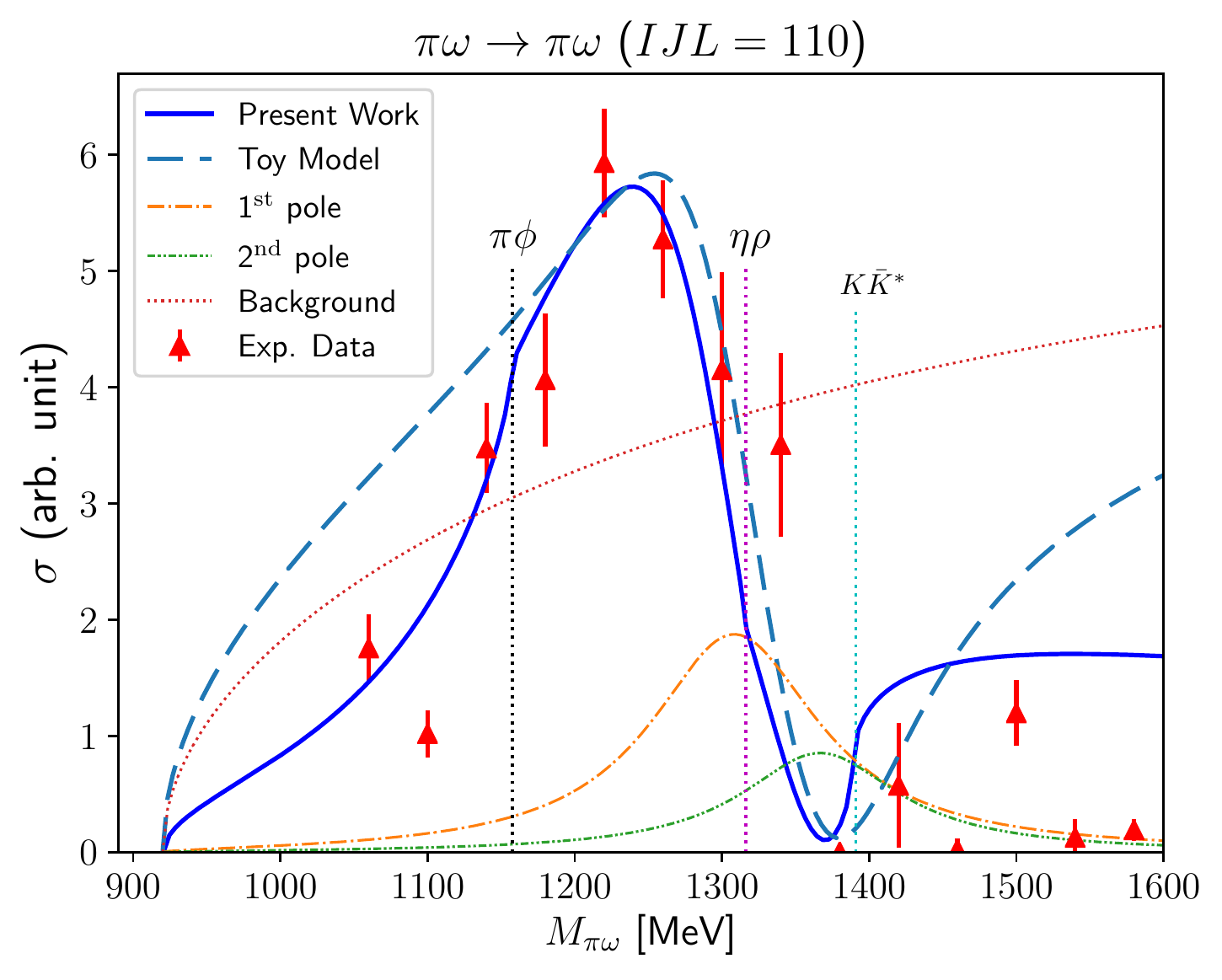}
 \caption{The $\pi\omega\to\pi\omega$ total cross
   section for $IJL=110$ as a function of $\pi\omega$ invariant
   mass. The experimental data are taken from
   Ref.~\cite{Fukui:1990ki}.}  
 \label{fig:6}
\end{figure}
To get a clearer picture of the two-pole structure of the $b_1$ meson 
and the effect of the phase of the coupling strengths, we propose a toy 
model that includes the two poles in the Breit-Wigner form. To describe 
the experimental data~\cite{Fukui:1990ki}, we need to introduce the 
background amplitude. So, the $T$ amplitude in the toy model is 
constructed as follows:
\begin{align}
	T(s) = T_\mathrm{pole}(s)+T_\mathrm{back}(s)
\label{eq:27}
\end{align}
where the pole part of the $T$ amplitude contains two resonances 
corresponding to those found in the current work. They are 
parametrized by the Breit-Wigner form
\begin{align}
	\frac{g_{\pi\omega b_1}^2}{m^2-s-im\Gamma }
\end{align}
where the masses, widths, and coupling strengths to the $\pi\omega$
channel are taken from Table~\ref{tab:2}. The $T_\mathrm{back}$ stands
for the constant background $T$ amplitude, which is introduced to fit
the data. The total cross section is then obtained as 
\begin{align}
	\sigma(s) = f \rho(s) |T(s)|^2 = f\frac{|\bm{p}|}{8\pi\sqrt{s}}
	|T(s)|^2
\end{align}
where $\bm{p}$ is the momentum of $\omega$ in the CM frame and the 
factor $f$ is introduced to fit the experimental data. In
Fig.~\ref{fig:6}, we draw the results from the toy model. 
The dot-dashed and dot-dot-dashed curves depict the first and second
resonances for the $b_1$ mesons, respectively, whereas the dotted one
displays the background contribution. Note that the $M_{\pi\omega}$
dependence of the background contribution arises from the phase-space
factor. The long dashed curve shows the total result given by
Eq.~\eqref{eq:27}, which describes the experimental data in the
vicinity of the resonance region relatively well. Note that it
deviates from the data below the $\pi\phi$ and above the $K\bar{K}^*$
thresholds, because the toy model aims only at 
the resonance region. It implies that the two poles can describe the
apparent resonance of the known $b_1(1235)$ meson. The $b_1(1235)$
arises from the constructive interference of the first pole and the
background contribution. On the other hand, the threshold behavior is
governed by the destructive interference of all the contributions. In
particular, the second pole comes into essential play in explaining
the threshold behavior. 

Actually, the two-pole structure in hadron spectroscopy is not a novel 
observation~\cite{Meissner:2020khl}. For example, the $\Lambda(1405)$
appears from the two $\Lambda$ poles, i.e. $\Lambda(1380)$ and
$\Lambda(1405)$~\cite{PDG}, which was initially found in
Ref.~\cite{Oller:2000fj}, where the $K^- p$ interaction was examined
in a chiral unitary approach~\cite{Oller:2000fj} with the
$\pi\Sigma$, $\eta\Lambda$ and $K\Xi$ channels coupled. A further
study~\cite{Jido:2003cb}  demonstrated that, owing to the phase
difference between the two poles, only a single resonance was 
observed experimentally. In addition, the same approach uncovered the
two-pole structure in the heavy meson
domain~\cite{Kolomeitsev:2003ac}: the excited $D$ and $B$ mesons with
quantum numbers $J^P=0^+$ and $1^+$ originate from the two-pole
resonances. The $K_1$(1270) meson in the light-meson
sector were also considered to have the two-pole
structures~\cite{Roca:2005nm,Geng:2006yb}. Notably, the two-pole
structure cannot easily be separated by experiments, and thus
experimental observations always led to a single
resonance~\footnote{Note that $\Lambda(1380)$ and $\Lambda(1405)$
  could be distinguished by the proton
  electroproduction~\cite{Nam:2017yeg}.}. The $b_1(1235)$ may also be 
the case. In particular, the behavior of the total cross section near
the $K\bar{K}^*$ threshold holds the significant clue about the fact
that the $b_1(1235)$ originates from the two $b_1$ mesons, $b_1(1306)$
and $b_1(1356)$.  

Finally, we want to discuss the coupling strengths of the two $b_1$
mesons. The results are given respectively as
$g_{\pi\omega}=(3.07+i2.14)$ GeV and $g_{\pi\omega}=(0.68+i2.91)$ GeV
for the first and second $b_1$ mesons, as listed in
Table~\ref{tab:2}. In particular, the coupling strength of the second
pole has the lager imaginary part than its real one. This finding is
similar to the case of the $\Lambda$(1405)~\cite{Jido:2003cb}, where
the coupling strength to the $\pi\Sigma$ channel of the higher energy
pole of $\Lambda$(1405) has the stronger imaginary part, resulting in
a threshold behavior near the $\bar{K}N$ threshold in $\pi\Sigma$ 
elastic scattering.

\section{Summary and conclusion}
In the current work, we aimed at investigating the $b_1$ meson in 
$\pi\omega$ scattering within the framework of the fully off-mass-shell 
coupled-channel formalism. Since the $b_1$ meson has the four decay
modes, we considered the corresponding channels, i.e., $\pi\omega$,
$\eta\rho$, $K\bar{K}^*$, and $\pi\phi$. We first constructed the
tree-level kernel amplitudes with all possible transitions, using the
effective Lagrangians with flavor SU(3) symmetry and hidden local
gauge symmetry for the vector mesons. We introduced the form factors
at each vertex. Using the kernel amplitudes, we solved the fully
off-mass-shell Blankenbecler-Sugar coupled integral equation. Since we
are interested in the $b_1$ channel, we performed the partial-wave
expansion to select the partial wave with $I=1$, $J=1$, and $L=0$,
which corresponds to the $b_1$ channel. We compared the numerical
results for the total cross section with the experimental data. The
results describe the $b_1(1235)$ resonance very well. They also reveal
the threshold behavior in the vicinity of the $K\bar{K}^*$
channel. We scanned the transition amplitude for the $\pi\omega$
interaction in the complex energy plane, we found the positions of the
two singularities: the one is located below the $\eta\rho$ threshold,
whereas the other is positioned above the $\eta\rho$ and below the
$K\bar{K}^*$ thresholds. We also extracted the coupling strengths of
the $b_1$ coupled to the four different channels in both the $S$ and
$D$ waves. To analyze the two-pole structure, we constructed the toy
model that contains the pole amplitudes for the two $b_1$ mesons in
the Breit-Wigner form and the background contributions. We observed
that the $b_1(1235)$ resonance appears from the constructive
interference of the first $b_1$ and background contribution whereas
the threshold behavior near the $K\bar{K}^*$ threshold arises from the
destructive interference of all the contributions. Specifically, the
second pole plays a critical role in describing the threshold behavior
of the total cross section. We concluded that the $b_1(1235)$ meson  
occurs from the two $b_1$ mesons, i.e., $b_1(1306)$ and $b_1(1356)$. 

\begin{acknowledgments}
We are grateful to Jung-Keun Ahn, Tetsuo Hyodo, Hee-Jin Kim, and
Seung-il Nam for fruitful discussions.  The present work was supported
by Inha University Grant, 2023.  
\end{acknowledgments}

\bibliography{b1meson}
\bibliographystyle{apsrev4-2}

\end{document}